\documentclass[a4paper,fleqn,usenatbib]{mnras}

\usepackage[T1]{fontenc}
\usepackage{ae,aecompl}
\usepackage[utf8]{inputenc}

\usepackage{graphicx}	
\usepackage{amsmath}	
\usepackage{amssymb}	
\usepackage{lscape}
\usepackage{tikz}   
\usepackage{textcomp}
\usepackage{gensymb}
\usepackage{tikz-3dplot} 
\usepackage{import}
\usepackage{lmodern}
\RequirePackage{fix-cm}

\usepackage{graphicx}	
\usepackage{amsmath}	
\usepackage{amssymb}	

\newcommand{\nobs}{\mathbf{n}_{\rm obs}}

\newcommand{\ex}{\mathbf{e}_{\rm x}}
\newcommand{\ey}{\mathbf{e}_{\rm y}}
\newcommand{\ez}{\mathbf{e}_{\rm z}}

\newcommand{\rlight}{r_{\rm L}}

\newcommand{\Rs}{R_{\rm s}}

\title[Radio and X-ray modelling of pulsars]{Joint radio and X-ray modelling of PSR~J1136+1551}

\author[J\'er\^ome P\'etri, Dipanjan Mitra]{
J. P\'etri,$^{1}$\thanks{E-mail: jerome.petri@astro.unistra.fr}
D. Mitra$^{2,3}$
\\
$^{1}$Universit\'e de Strasbourg, CNRS, Observatoire astronomique de Strasbourg, UMR 7550, F-67000 Strasbourg, France.\\
$^{2}$ National Centre for Radio Astrophysics, Tata Institute for Fundamental Research, Post Bag 3, Ganeshkhind, Pune 411007, INDIA\\
$3$Janusz Gil Institute of Astronomy, University of Zielona G\'ora, ul. Szafrana 2, 65-516 Zielona G\'ora, Poland
}

\date{Accepted XXX. Received YYY; in original form ZZZ}

\pubyear{2019}

\begin{document}
\label{firstpage}
\pagerange{\pageref{firstpage}--\pageref{lastpage}}
\maketitle

\begin{abstract}
Multi-wavelength observations of pulsar emission properties are powerful means to constrain their magnetospheric activity and magnetic topology. Usually a star centred magnetic dipole model is invoked to explain the main characteristics of this radiation. However in some particular pulsars where observational constraints exist, such simplified models are unable to predict salient features of their multi-wavelength emission. This paper aims to carefully model the radio and X-ray emission of PSR~J1136+1551 with an off-centred magnetic dipole to reconcile both wavelength measurements. We simultaneously fit the radio pulse profile with its polarization and the thermal X-ray emission from the polar cap hot spots of PSR~J1136+1551. We are able to pin down the parameters of the non-dipolar geometry (which we have assumed to be an offset dipole) and the viewing angle, meanwhile accounting for the time lag between X-rays and radio emission. 
Our model fits the data if the off-centred magnetic dipole lies about 20\% below the neutron star surface. We also expect very asymmetric polar cap shapes and sizes, implying non antipodal and non identical thermal emission from the hot spots. We conclude that a non-dipolar surface magnetic field is an essential feature to explain the multi-wavelength aspects of PSR~J1136+1551 and other similar pulsars.
\end{abstract}




\section{Introduction}

%

Rotation powered pulsars emits broadband electromagnetic radiation, due to relativistic particles streaming along open magnetic field lines in the magnetosphere, and the pulsed emission is seen across the spectrum. 
PSR~J1136+1551 is a middle aged, so called normal pulsar (with pulsar periods~$P$ longer than $\sim$~100~msec) with period $P = 1.19$~sec, and is seen to emit both in the radio and X-ray wavelength. The radio emission is coherent in nature and well constrained to originate close to the neutron star, typically below~10\% of the light cylinder. The X-ray emission comprises of primarily two sources, 
the thermal X-ray from hot polar cap that arises due to bombardment of 
back streaming particles on the neutron star surface, and the non-thermal X-ray 
whose origin is not well known and can arise due to acceleration of charge particles 
along the open magnetic field lines and/or inverse Compton processes in the magnetosphere.
Typically the non-thermal and thermal emission dominates at different parts of the X-ray spectrum. Combined model of thermal and non-thermal fits to the X-ray spectrum data is usually attempted to constrain
features like temperature and area of the thermal hotspot emission and obtain a power law index for the non-thermal emission.  For a pure black body emission, the  estimated hotspot area ($A_{\rm h}$) in fact corresponds to the geometrical area of the polar cap.  
Since for a given pulsar period, the theoretical polar cap 
area $A_{\rm d}$ for a star centred global dipole is known, it is
useful to compare $A_{\rm d}$ with $A_{\rm h}$, where $A_{\rm h}/A_{\rm d} \sim 1$ correspond
to a surface dipole magnetic field, and $A_{\rm d} / A_{\rm h} > 1$ correspond to multipolar 
magnetic field. To find the area $A_{\rm h}$, X-ray observation and spectral modelling of PSR~J1136+1551 
has been attempted by \cite{kargaltsev_x-ray_2006-1} and \cite{szary_xmm-newton_2017} S17 hereafter.
S17 work was a substantial improvement over \citet{kargaltsev_x-ray_2006-1} 
in terms to improving the X-ray photon statistics significantly,
and their combined fit to the data with a black body (BB) + power law (PL)  yielded 
$A_{\rm d}/A_{\rm h} > 1$, which led S17 to suggest the presence of surface multipolar
magnetic fields.

Unfortunately the above method of X-ray spectral fitting 
to obtain $A_{\rm h}$ from BB has several drawbacks, see e.g. \cite{arumugasamy_evaluating_2019}. 
Firstly for most pulsars the X-ray statistics is poor and hence it is difficult
to distinguish between models of BB or PL or BB+PL. For example 
in the case of PSR~J1136+1551  S17 found that all the models
fits the spectra with reasonable significance, and it is difficult
to find a preferred model.  Secondly there are several physical effects
that can reprocess the BB emission, like the presence of neutron 
star atmosphere or inverse Compton scattering of the BB due to 
back streaming particles, and hence the estimated $A_{\rm h}$ most likely does 
not correspond to the actual surface area. Thus, the conclusion that 
surface magnetic fields are multipolar in nature based on X-ray spectral
fits are inconclusive and uncertain.  

S17 also checked for time alignments between radio and X-ray profiles 
for PSR~J1136+1551 by dividing the X-ray spectra in several 
energy ranges: 0.2--0.5 keV, 0.5--1.2 keV, 1.2--3 keV, and 0.2--3 keV, and  
found the light curves to have an offset (called X-R offset hereafter) 
of $70^{\circ}\pm 8^{\circ}$, $44^{\circ}\pm 9^{\circ}$, $92\pm7^{\circ}$ 
respectively.  Generally the lower energy ranges in the X-ray spectrum 
is BB dominated while the higher energy is PL dominated. However
this aspect cannot be resolved for PSR~J1136+1551, and hence the
X-R offset at the least suggest that
the radio emission leads the X-ray emission 
by about $64^{\circ}\pm 7 ^{\circ}$, where the X-ray emission can have 
contribution from thermal or non-thermal or a combination of both. S17 first considered
the X-R offset to arise due to surface thermal X-ray and 
radio emission arising from a few hundred km above the neutron star surface. 
In this case to explain the offset, S17 made rough estimates for the 
displacement of the polar cap to be
about 9.7 km from the neutron star centre, which is almost the neutron star radius.
Stating that such large displacements are not physically justifiable, S17
suggested that the X-R offset is possibly arising due to non-thermal X-ray.

In this work we revisit the problem of how to explain the X-R offset in a significantly
more quantitative manner than has been attempted before. Since the 
X-ray observations cannot be used to disentangle the thermal and non-thermal  
emission, we will consider both the possibilities. Our work benefits from 
several important recent theoretical developments that allow us to study
the pulsar magnetosphere in a quantitative manner. Indeed, force-free pulsar magnetospheres can now be computed accurately in full 3D geometry (\citealt{spitkovsky_time-dependent_2006}, \citealt{petri_pulsar_2012}). Moreover, there are some hints for the presence of non-dipolar surface magnetic fields. The simplest approach is to take an off-centred dipole as done by \cite{petri_radiation_2016} who also computed the expected polarization signature in \cite{petri_polarized_2017}. In this last work, \cite{petri_polarized_2017} already claimed that X-R offset can be explained by the off-centred dipole. To support our idea, we model the radio and X-ray emission from PSR~J1136+1551 for which good data sets are available.

The organization of the paper is as follows. In section~\ref{sec:radio_obs} we  discuss the methods of finding the radio emission geometry and location of the radio emission regions for PSR~J1136+1551. We verify the validity of these methods by comparing it with predictions of various models of the pulsar magnetosphere.
In section~\ref{sec:x-r_offset}, we use a simple model of an offset dipole to estimate the observed X-R offset, in both the thermal and non-thermal case. In section~\ref{sec:psr_j1136} we apply our results to PSR~J1136+1551. A discussion on the possibility of non-thermal X-ray emission is discussed in section~\ref{sec:non_thermal_emission} before concluding by section~\ref{sec:conclusion}.

\section{Radio Observations, polarization and Emission heights}
\label{sec:radio_obs}

The full polarimetric radio observations can be used to make estimates 
of the dipolar emission geometry at the radio emission region for PSR~J1136+1551. 
For this purpose we use archival average full polarization pulsar data obtained 
from the Giant Meter-wave Radio Telescope at 339~MHz and 618~MHz respectively for
the Meter-wavelength Single pulse polarimetric survey (MSPES, \cite{mitra_meterwavelength_2016}).
The 618~MHz data is published, and the 339~MHz data was a part
of the test data taken during the MSPES.

\begin{figure}
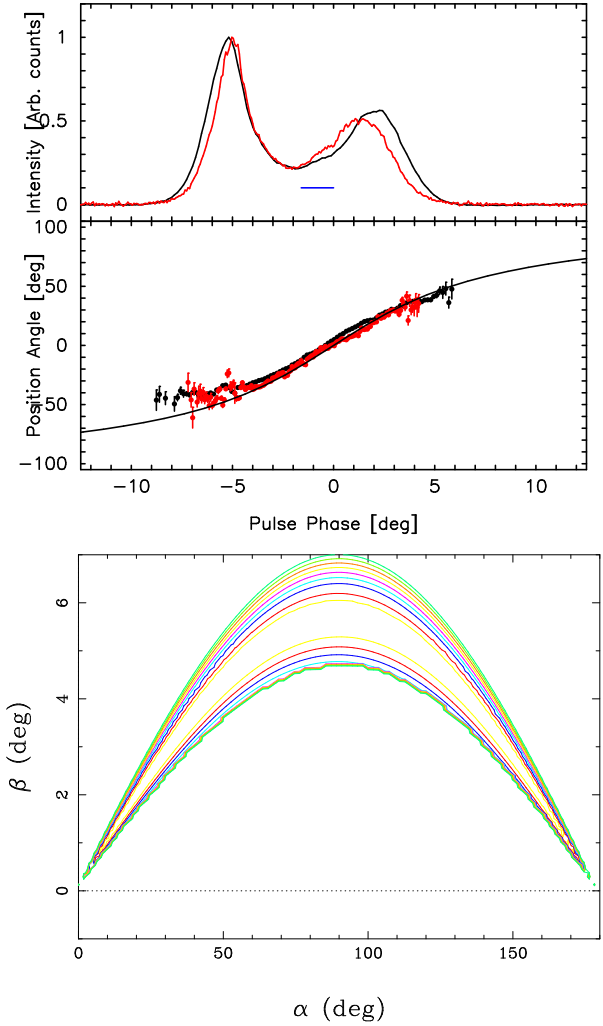

	\begin{center}
		\begin{tabular}{cc}
			\mbox{\includegraphics[width=0.4\textwidth,angle=-90.]{combined.ps}}\\
			\mbox{\includegraphics[width=0.35\textwidth,angle=-90.]{chi2_300_610.ps}}\\
		\end{tabular}
		\caption{Top plot shows the average profile of PSR~J1136+1551
			at 339~MHz (black) and 618~MHz (red). The top panel shows the total intensity
			profile, and the point in the bottom panel is the PPA. The solid line 
			displays the RVM fit using E.~(\ref{req1}) for which $\alpha = 130$\degr 
			and $\beta=4.2$\degr. The bottom plot shows the $\chi^2$ distribution for 
                        the fitted parameter $\alpha$ and $\beta$, where we clearly see that the 
                        parameters are highly correlated.}
		\label{fig1}
	\end{center}
\end{figure}

Given the full polarization data, the first step  is to access the validity of the rotating vector model (RVM
hereafter) proposed by \citet{radhakrishnan_magnetic_1969}. 
According to the RVM, the linear polarization vectors are modelled
to be parallel to the projection of the magnetic field line in the plane
of the sky. As the star rotates, the line of sight traverses the emission region, 
the angle ($\Psi$) made by the projected vectors changes as a 
function of pulse phase ($\phi$).
For a star centred dipolar magnetic field if $\alpha$ is 
the angle between the rotation axis and the magnetic axis, and 
$\beta$ is the impact angle, then, introducing the inclination angle~$\zeta=\alpha+\beta$ between the line of sight and the rotation axis,
the RVM has a characteristic S-shaped traverse given by, 
\begin{equation}
\Psi = \Psi_{\circ} + \tan^{-1} \left( \frac{\sin\alpha\sin(\phi-\phi_{\circ})}
{\sin\zeta\cos\alpha - \sin\alpha \cos \zeta \cos(\phi-\phi_{\circ})}\right)
\label{req1}
\end{equation}
Here $\Psi_{\circ}$ and $\phi_{\circ}$ are the arbitrary phase offsets
for the polarization angle~$\Psi$ and phase~$\phi$ respectively. At $\Psi_{\circ}$ the polarization position angle (PPA) goes through the steepest gradient (SG) point, which for a static dipole 
magnetic field is associated with the plane containing the rotation and 
the magnetic axis.  We fit Eq.~(\ref{req1}) to  
the polarization data of PSR~J1136+1551 at both 339 and 618~MHz respectively, and find 
that the RVM is a very reasonable model. This is consistent with the finding of \citet{2004A&A...421..215M},
that in pulsar the shape of the PPA traverse is frequency independent, and 
further we use their method for combining the PPA at two frequencies. To do this
we first fit the RVM to get $\Psi_{\circ}$ and $\phi_{\circ}$ at each frequencies.
Then we subtract the offsets and to obtain the combine PPA, as shown in
top plot bottom panel of Fig.~\ref{fig1}. We now use this combined PPA
and fit the RVM to obtain $\alpha$ and $\beta$, with the offsets being
set to zero. Although in most cases the RVM fit to the PPA traverse is acceptable, the estimates of the geometrical angles $\alpha$ and $\beta$ are highly correlated, as has been also shown by a large number of studies (\citealt{1997A&A...324..981V}, \citealt{2001ApJ...553..341E}, \citealt{2004A&A...421..215M}). This is mostly due to the fact that significantly wider profiles than mostly observed
are needed to  distinguish the geometrical angles using RVM. 
For the combined PPA traverse we fit the RVM  using Eq.(\ref{req1}) and
also find the $\alpha$ and $\beta$ values to be highly correlated as shown
in the $\chi^2$ contour plot in the bottom panel of Fig.~\ref{fig1}. 

In the top plot (bottom panel) the RVM fit (black line) is shown for parameters $\alpha = 130^{\circ}\pm 10^{\circ}$ and $\beta = 4.2^{\circ} \pm 0.5^{\circ}$. The choice of $\alpha$ and $\beta$ is somewhat arbitrary, but we will justify below our preference for these values. Note that the RVM (back line) goes below the data points around -5$^{\circ}$ longitude, and this is due to the fact that the average PPA is affected due to the presence of orthogonal polarization moding which can be clearly seen in single pulse observations.  The phase offsets have been subtracted and have errors of $\phi_{\circ} = 0.0^{\circ} \pm 0.5^{\circ}$ and $\Psi_{\circ} = 0.0^{\circ} \pm 5^{\circ}$. 
Clearly it is futile to get realistic estimates of the geometrical angles using
the RVM. However, the SG point is related to the geometrical angles as 
\begin{equation}
\frac{\sin \alpha}{\sin \beta} = \left| \frac{d\Psi}{d\phi} \right|_{\rm max} .
\label{req2}
\end{equation}
From  the RVM fits, generally the location of the phase of the steepest gradient point,  $\phi_{\circ}$ 
is significantly better constrained, and we find that for PSR~J1136+1551, 
$\mid d\Psi/d\phi \mid_{\rm max} = 10.5\pm2$. 

Since the geometry cannot be constrained by RVM fits, the Empirical Theory (ET) of Pulsar emission (\citealt{1983ApJ...274..333R}, ETI; \citealt{1993ApJ...405..285R},ETVI; \citealt{2002ApJ...577..322M},ETVII) provides an alternative. In ETVI it was proposed that the two dimensional pulsar radio emission beam at 1~GHz is circular in shape and is organized in the form of a central core emission with two nested, so called inner and outer conal emission structure. Assuming spherical geometry, the radius of the emission beam $\rho$ is connected to $\alpha$, $\beta$ and the width of the profile $W$ as,
\begin{equation}
\sin^2(\rho/2) = \sin(\alpha + \beta)\sin(\alpha)\sin^2(W/4)+\sin^2(\beta/2).
\label{req3}
\end{equation} 
Depending on the line of sight of the observer, different number of components are seen in the
pulse profile. This gave rise to the classification scheme where profiles with five or three components are called Multiple (M) or Triple (T) class, and they correspond to central 
cuts of the beam with steep PPA traverses. For more tangential line
of sight with shallow PPA traverses, one of two component profile is seen
which are known as conal single (S$_{\rm d}$) and conal double (D) profiles.
In ETVI it was established that the inner and outer conal beam radii
$\rho^{\rm 1\,GHz}_{\rm inner/outer}$ measured at 1 GHz for various pulsars follow a straightforward scaling
relation with pulsar period, as 
\begin{equation}\label{eq:Width}
\rho^{\rm 1GHz}_{\rm inner/outer} = 4.3^{\circ} / 5.7^{\circ} P^{-0.5}.
\end{equation}

Also in ETVII it was shown that the outer conal components follow the phenomenon of radius to frequency 
mapping (RFM) where the pulse widths measured at outer half-power
points decreases with increasing frequency, where as for the inner
components the width tends to remain constant across frequency.  

The above ideas have been thoroughly applied to PSR~J1136+1551 and 
a detailed analysis of profile classification carried out in
ETVI positioned the pulsar to be D-type. 
In ETVII it was shown that
PSR~J1136+1551 outer component width follow the RFM property of that
of an outer conal component and hence $\rho^{\rm 1\,GHz}_{\rm outer} = 5.2^{\circ}$ 
(since $P=1.19$ sec).  
This fact is also corroborated by the detailed single
pulse analysis of PSR~J1136+1551 by \cite{2012MNRAS.424.2477Y}, 
where they show evidence for the existence of both inner and 
outer conal component.   Now knowing the measured width of 
the pulsar $W_{\rm 1\,GHz}$ at 1~GHz, we can use Eq.~(\ref{req3})
to find the pulsar geometry.  In Eq.~(\ref{req3}) we know $\rho$ and 
$\beta$ can be written in terms of $\alpha$ using Eq.~(\ref{req2})
and further we can now use an iterative procedure to find appropriate
$\alpha$ and $\beta$ that will yield values of width $W_{\rm 1\,GHz}$ that agrees
with the observed value. The measured width at 1~GHz at the outer half-power 
point $W_{\rm 1\,GHz} = 8.5^{\circ}\pm 0.4^{\circ}$, and this width can be fitted well with 
$\alpha = 130^{\circ}$ and $\beta = 4.2^{\circ}$. By definition this positive value of $\beta$ obtained for the case $\alpha > 90^{\circ}$ corresponds to the so called inner line of sight geometry. Note that the outer line of sight solution is $\alpha = 50^{\circ}$ and $\beta = 4.2^{\circ}$ works
as well as the inner line of sight solution for the given, since 
the effect of inner and outer is only seen in wide profile widths. However,
as we will justify later, in this work we have the preference for the inner line of sight geometry. 
Assuming a star centred dipolar magnetic field and the emission across the
profile being generated from the same emission height (see ETVI) 
above the neutron star of radius 10~km, the radio emission height
can be computed as 
\begin{equation}
\label{eq:Altitude}
h = 10 \, P \, \left(\frac{\rho}{1.23}\right)^2 \textrm{km} \sim 214\ \textrm{km}.
\end{equation}
$P$ is expressed in seconds and $\rho$ in degrees.

\subsection{A/R Emission heights}

RVM assumes a static star centred dipole magnetic field. However
in reality the star is rotating and if the radio emission originates
at a height $h$ above the neutron star, then the effect of aberration
and retardation (A/R hereafter) needs to be included. Interestingly,
as shown by several studies (\citealt{1991ApJ...370..643B}, \citealt{2008MNRAS.391..859D}, 
\citealt{2001ApJ...546..382H}, \citealt{lyutikov_relativistic_2016}), there is an observational effect associated
with the A/R effect, where the phase at the center of the observed pulse profile
leads the SG point of the PPA traverse by an angle $\Delta \phi_{\rm obs}$ degrees. 
For slowly rotating normal pulsars, and emission arising below 10\% of
the light cylinder, the linear approximation of the A/R effect 
can be used, where $\Delta \phi_{\rm obs}$ is related to 
emission height as 
\begin{equation}\label{eq:hAR}
h_{\rm A/R} = \frac{c \, P \, \Delta \phi_{\rm obs}}{1440}\ \textrm{km}.
\end{equation}
where $c$ is the velocity of light.

For PSR~J1136+1551 we measure the midway point of the profile center based on the outer 10\% widths of the total intensity pulse profile and find that for both 339 and 618~MHz the point leads the SG point, i.e. $\Delta \phi_{\rm obs} = -1.6\degr \pm 0.1\degr$. In Fig.\ref{fig1} the length $\Delta\phi_{\rm obs}$ is shown as a blue line in the top plot. The corresponding altitude is $h_{\rm A/R} \sim 393 \pm 25$ km.

\subsection{Validity of the A/R method} 

The A/R shift of PPA with respect to the pulse profile centre relies mainly on a centred static magnetic dipole and vacuum field in the magnetosphere. However in reality the magnetosphere is filled with plasma and is best described by presence of non-dipolar surface magnetic field and the Deutsch solution \cite{deutsch_electromagnetic_1955}. And all these effects can in principal influence the estimate of the radio emission height as given by Eq.~(\ref{eq:hAR}). 

In this section, we carefully quantify the shift introduced by these supplementary effects by considering various conditions of the magnetosphere, and a rotating off centred magnetic dipole as a model for non-dipolar magnetic field which has been developed in \citep{petri_radiation_2016,2017MNRAS.466L..73P}
and is also described in section~\ref{sec:x-r_offset}.
Analytical expressions derived for the vacuum field can then be compared with our numerical treatment. 

Let us briefly review the different configurations accounting for A/R effects. For emission 
arising at a height $r$ and the light cylinder distance $\rlight=c\,P/2\pi$, aberration 
leads to a first order delay in time of arrival such that according to \cite{2004ApJ...614..869D}
\begin{equation}
\label{eq:AberrationDelai}
\Delta \phi_{\rm ab} = - \frac{r}{\rlight} .
\end{equation}
Retardation leads to another time delay of the same order of magnitude, contributing in the same direction, i.e. a delay (with a minus sign), such that
\begin{equation}
\label{eq:RetardDelai}
\Delta \phi_{\rm ret} = - \frac{r}{\rlight}
\end{equation}
both depending linearly on the emission height~$r$. These estimates are geometry independent, therefore very robust. As an additional geometry dependent effect, magnetic field sweep back due to rotation tries to cancel these effects in such a way that 
\cite{shitov_period_1983}
\begin{equation}\label{eq:SweepBack}
\Delta \phi_{\rm sb} \approx 1.2 \, \left( \frac{r}{\rlight} \right)^3 \, \sin^2 \alpha
\end{equation}
which is negligible well inside the light-cylinder, compared to the former delays. A much more important perturbation is related to the global shift of the polar cap centre with respect to the magnetic poles. The displacement of the polar cap rims produces another shift in the opposite direction to aberration and retardation, and equal to
\begin{equation}\label{eq:CapDelai}
\Delta \phi_{\rm ov} \approx 0.2 \, \sqrt{\frac{r}{\rlight}} \sim r^{0.5}
\end{equation}
which is of half-order $0.5$ in emission height exponent. It is the dominant effect for very low emission altitudes 
\citep{dyks_rotational_2004}. 
Note also that the polar caps are defined by the global magnetospheric structure, not only by considering locally electrodynamics close to the surface. 

\begin{figure}
	\includegraphics[width=0.5\textwidth]{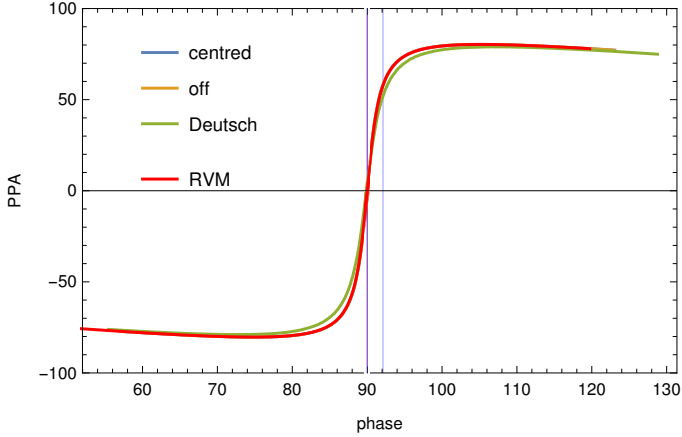}
	\caption{PPA and its inflexion point compared to the size of the polar cap in several approximations: a static centred dipole, a static off-centred dipole and the Deutsch solution. The RVM is shown in red for reference. No A/R effects are included. $\alpha=50\degr$ and $h/\rlight=0.08$.}
	\label{fig:AR_comparison_RVM}
\end{figure}

All these contributions have a strong impact on the shift between the middle of the radio pulse profile and the PPA inflexion point. We quantify precisely these effects by numerical simulations taking into account a rotating dipole or an off-centred dipole. The neutron star spin is equal to $P=1.19$~s corresponding to PSR~J1136+1551. First, in Fig.~\ref{fig:AR_comparison_RVM} we show the PPA in the RVM model in red solid line and compare it to the centred dipole in blue, the off-centred dipole in orange, and the Deutsch field in green. All PPA are undistinguishable when emission emanates well inside the light-cylinder. Thus the inflexion point is the same, depicted by a orange vertical bar around a phase $\phi=90\degr$. What is affected by these models is the location of the polar cap rim. For the static dipole, it is centred around phase $\phi=90\degr$, thus no shift between pulse profile and PPA. For the off-centred dipole, the trailing part of the pulse is shorter, shifting the middle of pulse profile to slightly earlier phases with respect to PPA. Finally, for the Deutsch solution, the polar cap size is much larger, the leading side being increase by~$2\degr$ whereas the trailing side being increased by~$5\degr$. This causes a net shift at later phases compared to PPA, as predicted by 
\cite{dyks_rotational_2004}. 
The blue vertical bar shows the location of the pulse profile centre in the different approximations. Note that the polar cap rim deducted from the magnetic field sweep back contributes oppositely to A/R effects.

\begin{figure}
	\includegraphics[width=0.5\textwidth]{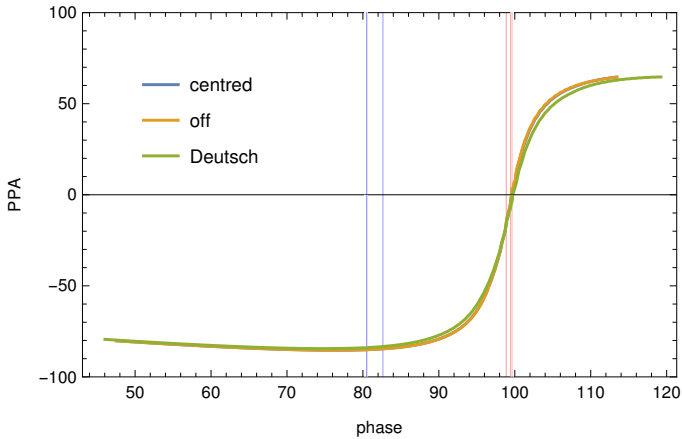}
	\caption{PPA and its inflexion point compared to the size of the polar cap in several approximations: a static centred dipole, a static off-centred dipole and the Deutsch solution. A/R effects are included. $\alpha=50\degr$ and $h/\rlight=0.08$.}
	\label{fig:AR_comparison}
\end{figure}

Next we add A/R effects to the geometry. The new PPA and pulse profile sizes are shown in Fig.~\ref{fig:AR_comparison}. The PPA inflexion point is located around $100\degr$ in all cases but the middle of the pulse profile is around $81\degr-83\degr$. The Deutsch field counterbalances the A/R effects by reducing the shift as seen in this plot by computing the distance between the blue vertical line and the orange vertical line. 

\begin{figure}
	\includegraphics[width=0.5\textwidth]{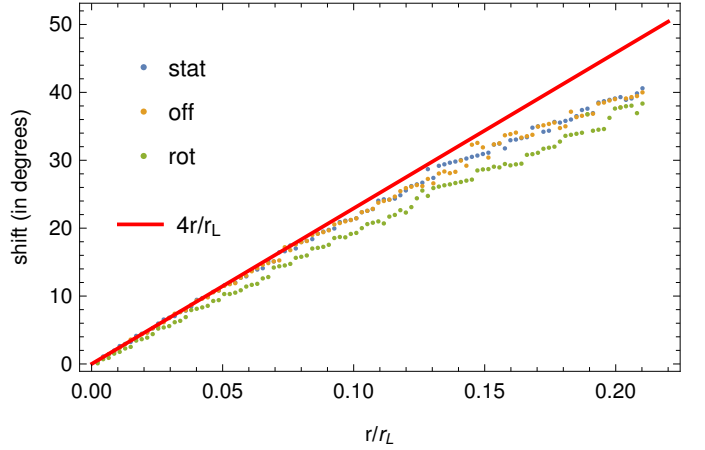}
	\caption{Evolution of the shift between PPA inflexion point and pulse profile centre in several approximations: a centred dipole, an off-centred dipole and the Deutsch solution with $\alpha=50\degr$ and $\beta=1$\degr. The standard expectation is shown in red for reference.}
	\label{fig:AR_deviation_alpha50_beta1}
\end{figure}

\begin{figure}
	\includegraphics[width=0.5\textwidth]{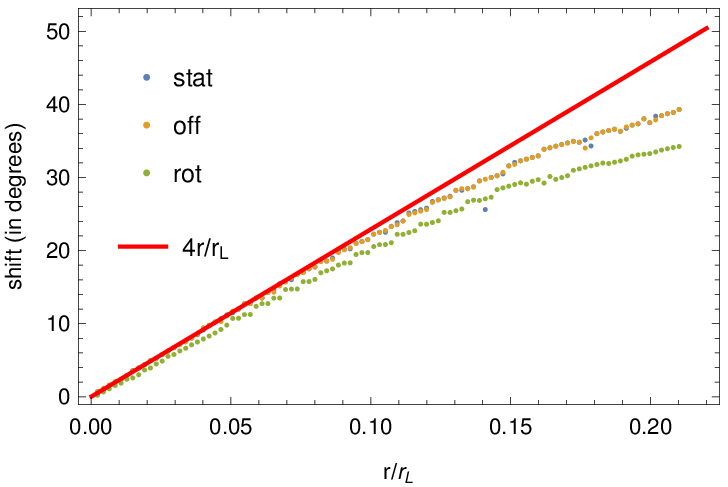}
	\caption{Evolution of the shift between PPA inflexion point and pulse profile centre in several approximations: a centred dipole, an off-centred dipole and the Deutsch solution with $\alpha=90\degr$ and $\beta=1$\degr. The standard expectation is shown in red for reference.}
	\label{fig:AR_deviation_alpha90_beta1}
\end{figure}

\begin{figure}
	\includegraphics[width=0.5\textwidth]{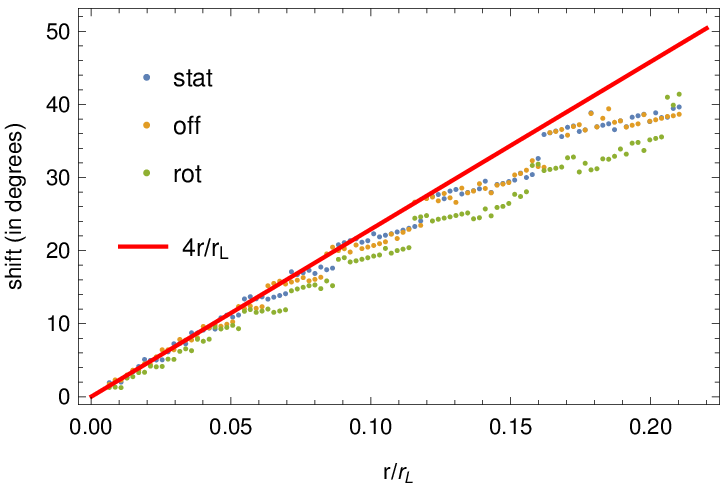}
	\caption{Evolution of the shift between PPA inflexion point and pulse profile centre in several approximations: a centred dipole, an off-centred dipole and the Deutsch solution with $\alpha=50\degr$ and $\beta=5$\degr. The standard expectation is shown in red for reference.}
	\label{fig:AR_deviation_alpha50_beta5}
\end{figure}

\begin{figure}
	\includegraphics[width=0.5\textwidth]{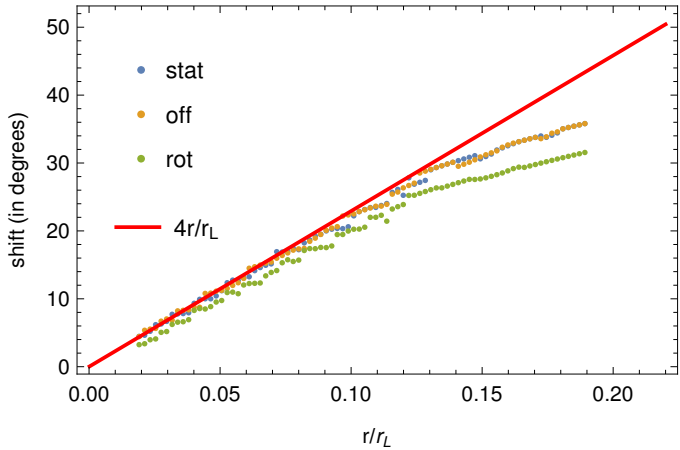}
	\caption{Evolution of the shift between PPA inflexion point and pulse profile centre in several approximations: a centred dipole, an off-centred dipole and the Deutsch solution with $\alpha=90\degr$ and $\beta=5$\degr. The standard expectation is shown in red for reference.}
	\label{fig:AR_deviation_alpha90_beta5}
\end{figure}

The A/R effects are usually summarized by a simple formula given by Eq.~(\ref{eq:hAR}), which in terms of shift $\Delta \phi$ can be written
as 
\begin{equation}\label{eq:ARformula}
\Delta \phi \approx 4\,r/\rlight.
\end{equation}
In order to check its validity with emission height, we plot the measured shift and the expectations for several geometries and a bunch of emission heights. Results are summarized in Fig.~\ref{fig:AR_deviation_alpha50_beta1} for $\alpha=50\degr$ and $\beta=1\degr$, in Fig.~\ref{fig:AR_deviation_alpha50_beta5} for $\alpha=50\degr$ and $\beta=5\degr$, in  Fig.~\ref{fig:AR_deviation_alpha90_beta1} for $\alpha=90\degr$ and $\beta=1\degr$ and in  Fig.~\ref{fig:AR_deviation_alpha90_beta5} for $\alpha=90\degr$ and $\beta=5\degr$. The evolution of the A/R shift with distance is clearly seen according to the three magnetic field models. Generally, we notice that the analytical approximation $4\,r/\rlight$ is satisfactory up to 10\% of the light cylinder although it is systematically overestimated especially for the Deutsch solution.

We have therefore shown that the A/R formula is a very robust tool to estimate radio emission heights, whatever the geometry of the magnetic field close to the surface, dipolar or non dipolar. Radio emission probes the dipolar structure of the magnetosphere at about 10\% of the light-cylinder~$\rlight$. In this region, for normal pulsars, on one side, the emission height is large compared to the neutron star radius, therefore the multipolar components already decrease and become negligible (see also 
\cite{2002A&A...388..235G}), 
on the other side, the emission altitude remains well within the light-cylinder. Consequently, magnetic field distortion by magnetosperic current or retardation effect due to the finite speed of light remains small. However, our results would fail for millisecond pulsars where emission altitudes are only several neutron star radii, meanwhile close to the light-cylinder.

\section{Rotating off-centred dipole} 
\label{sec:x-r_offset}

We consider a simple off-centred magnetic dipole, introducing the relevant geometric parameters following the notation given by \cite{petri_radiation_2016} for a radiating dipole in vacuum with slight changes. 
For the emission processes, let it be synchrotron, curvature or inverse Compton, we neglect retardation effects as well as rotational sweep back of magnetic field lines. 

First we recall the important geometrical quantities and the magnetic configuration. Second we compute the polar cap distortion implied by the off-centering. Third, we derive an analytical formula for the time lag between thermal X-ray emanating from the hot spots and radio emission coming out from an altitude much less than the light-cylinder. Required vectors are expanded onto a Cartesian orthonormal basis $(\ex,\ey,\ez)$.

\subsection{Geometrical set-up}

\begin{figure}
\centering

\tdplotsetmaincoords{70}{110}

\pgfmathsetmacro{\rvec}{.6}
\pgfmathsetmacro{\thetavec}{60}
\pgfmathsetmacro{\phivec}{60}

\begin{tikzpicture}[scale=4,tdplot_main_coords]

\coordinate (O) at (0,0,0);

\tdplotsetcoord{P}{\rvec}{\thetavec}{\phivec}

\tdplotsetcoord{A}{1}{60}{0}

\tdplotsetcoord{B}{1.5}{60}{0}


\draw[thick,->] (0,0,0) node[below] {$O$} -- (1,0,0) node[anchor=north east]{$x$};
\draw[thick,->] (0,0,0) -- (0,1,0) node[anchor=north west]{$y$};
\draw[thick,->] (0,0,0) -- (0,0,1) node[anchor=south]{$z$ (spin axis)};

\tdplotsetrotatedcoords{0}{0}{0}

\draw[-stealth,thick,color=red] (O) -- (P) node [midway, above] {$d$} node [above, left] {$M$} ;
\draw[dashed,color=red,tdplot_rotated_coords] (0,0,0) -- (0.26,0.45,0);
\draw[dashed,color=red,tdplot_rotated_coords] (0.26,0.45,0) -- (0.26,0.45,.3);
\tdplotdrawarc[tdplot_rotated_coords,color=red]{(0,0,0)}{0.4}{0}{60}{anchor=north west}{$\Omega\,t$}

\draw[thick,color=green] (O) -- (A) node [midway, below] {$R$} ;
\draw[<->,thick,color=green] (A) -- (B) node [midway, below] {$h$} node [above, right] {$A$} ;



\tdplotsetthetaplanecoords{\phivec}

\tdplotdrawarc[red, tdplot_rotated_coords]{(0,0,0)}{0.5}{0}{\thetavec}{anchor=south west}{$\delta$}

\tdplotsetthetaplanecoords{0}

\tdplotdrawarc[green, tdplot_rotated_coords]{(0,0,0)}{0.5}{0}{60}{anchor=south east}{$\zeta$}


\tdplotsetrotatedcoords{0}{0}{0}

\tdplotsetrotatedcoordsorigin{(P)}

\draw[thick,tdplot_rotated_coords,->] (0,0,0) -- (.5,0,0) node[anchor=north east]{$x'$};
\draw[thick,tdplot_rotated_coords,->] (0,0,0) -- (0,.5,0) node[anchor=west]{$y'$};
\draw[thick,tdplot_rotated_coords,->] (0,0,0) -- (0,0,.5) node[anchor=south]{$z'$};


\draw[-stealth,thick,color=blue,tdplot_rotated_coords] (0,0,0) -- (.2,.2,.2) node [right] {$\pmb{\mu}$} ;
\draw[dashed,color=blue,tdplot_rotated_coords] (0,0,0) -- (.2,.2,0);
\draw[dashed,color=blue,tdplot_rotated_coords] (.2,.2,0) -- (.2,.2,.2);

\tdplotdrawarc[tdplot_rotated_coords,color=blue]{(0,0,0)}{0.2}{0}{45}{anchor=north west}{$\gamma+\Omega\,t$}

\tdplotsetrotatedthetaplanecoords{45}

\tdplotdrawarc[tdplot_rotated_coords,color=blue]{(0,0,0)}{0.2}{0}{55}{anchor=south west}{$\alpha$}

\begin{scope}[canvas is xy plane at z=0]
     \draw (1,0) arc (0:90:1);
\end{scope}
\begin{scope}[canvas is xz plane at y=0]
     \draw (1,0) arc (0:90:1);
\end{scope}
\begin{scope}[canvas is yz plane at x=0]
     \draw (1,0) arc (0:90:1);
\end{scope}   
\end{tikzpicture}
\caption{Geometry of the decentred magnetic dipole showing the three important angles $\{\alpha, \gamma, \delta\}$ and the displacement~$d$. Two additional parameters related to observations are the line of sight inclination~$\zeta$ and the emission height~$h$. The plot corresponds to time~$t$ assuming that the magnetic moment~$\bmu$ lies in the $(xOz)$ plane at $t=0$.}
\label{fig:Dipole}
\end{figure}
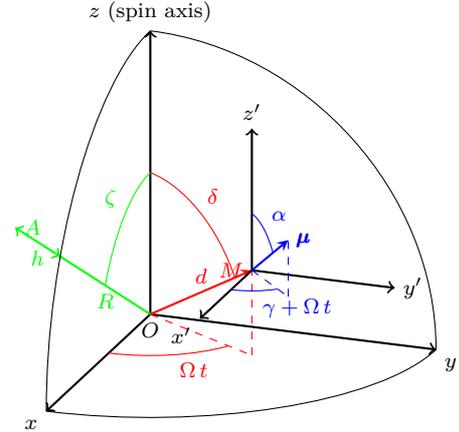

The neutron star is depicted as a solid body in uniform rotation at a rate~$\Omega$ along the $\ez$ axis. Its magnetic moment is located inside the sphere of radius~$R$ at a point~$M$ such that at any time $t$ its position vector is
\begin{equation}
\mathbf{d} = d \, (\sin \delta \, \cos \Omega\,t, \sin \delta \, \sin \Omega\,t, \cos \delta)
\end{equation}
where $d$ is the distance from the centre and $\delta$ the colatitude. Entrainment by the star is included in the phase term $\Omega\,t$. At the same time the magnetic moment~$\bmu$ points toward a direction depicted by the two angles $(\alpha,\gamma)$ and given by the unit vector
\begin{equation}
\mathbf{m} = (\sin \alpha \, \cos (\gamma+\Omega\,t), \sin \alpha \, \sin (\gamma+\Omega\,t), \cos \alpha) .
\end{equation}
The observer line of sight represented by the unit vector~$\nobs$ is by convention located at any time in the~$(xOz)$ plane, forming an angle~$\zeta$ with the spin axis ($\ez$ axis) thus 
\begin{equation}
\nobs = (\sin \zeta, 0, \cos \zeta) .
\end{equation}
The emission altitude, measured starting from the surface is denoted by~$h$. All important geometrical parameters are summarized in Fig.~\ref{fig:Dipole}.

The magnetic poles are defined by the intersection between the stellar surface, i.e. a sphere of radius~$R$, and the magnetic moment axis~$\bmu$. Their positions are found following the procedure we now describe. Let a sphere of radius~$R$ be centred at the origin of the reference frame. The intersection between this sphere and the straight line passing through the magnetic dipole moment located at~$M$ along its direction $\mathbf{m}$ is parametrized by a real parameter~$\lambda$ such that $\mathbf{r} = \lambda \, \mathbf{m} + \mathbf{d}$. We look for values of~$\lambda$ satisfying the relation $||\mathbf{r}|| = R$. This is equivalent to a quadratic equation in $\lambda$ requiring $\lambda^2 + 2 \, \lambda \, \mathbf{m} \cdot \mathbf{d} + d^2 - R^2 = 0$. The discriminant of this equation is equal to $\Delta = 4 \, ( (\mathbf{m} \cdot \mathbf{d})^2 + R^2 - d^2 )$ and always positive since $d<R$. Solutions are therefore always real and equal to
\begin{equation}
\label{eq:lambda}
 \lambda_\pm = -\mathbf{m} \cdot \mathbf{d} \pm \sqrt{(\mathbf{m} \cdot \mathbf{d})^2 + R^2 - d^2}
\end{equation}
with $\lambda_-<0$ and $\lambda_+>0$ and from which we deduce the poles at position
\begin{equation}
\label{eq:pole}
\mathbf{r}_\pm = \lambda_\pm \, \mathbf{m} + \mathbf{d}
\end{equation}
with 
\begin{equation}
\mathbf{m} \cdot \mathbf{d} = d \, ( \cos \alpha \, \cos \delta + \sin \alpha \, \sin \delta \, \cos \gamma ) .
\end{equation}
The positive solution~$\lambda_+$ is called the north pole whereas the negative solution~$\lambda_-$ is called the south pole.

The associated polarization angle~$\Psi$ has been found by \cite{petri_polarized_2017}. We call it decentred RVM (DRVM).
In this DRVM, contrary to the traditional RVM, the PPA depends on the emission height~$h$, conveniently normalized to the neutron star radius by~$\eta=h/R$, as well as on the displacement~$d$, also normalized to the neutron star radius according to~$\epsilon=d/R$. 
It represents the straightforward extension of the RVM \citep{radhakrishnan_magnetic_1969} for any displacement~$\epsilon\leq1$. The PPA is simply interpreted as the projection of the magnetic field line onto the plane of the sky when the system rotates. We emphasize that this polarization angle is now also impacted by the emission height~$h$ whenever~$d\neq0$. However, if the photons emanate from high altitudes compared to the stellar size, $h\gg R$, the DRVM reduces to the RVM within small corrections of the order $R/h = 1 /\eta \ll 1$. This means that the off-centred dipole as seen from large distances is undistinguishable from the centred dipole as long as the radio polarization is concerned. This situation is similar to a localized distribution of charges producing multipolar electric fields, perceptible close to the location of the source but tending to the lowest order multipole component being usually a monopole or a dipole. Therefore for high altitude emission $\eta\gg1$ we have
\begin{equation}\label{eq:DRVM}
\Psi_{\rm DRVM} = \Psi_{\rm RVM} + O\left( \frac{1}{\eta} \right)
\end{equation}
where $\Psi_{\rm RVM}$ is given by Eq.~(\ref{req1}). Note that for DRVM, the line of sight inclination~$\zeta$ is different from $\alpha+\beta$. However for high altitudes $\eta\gg1$, we also have $\zeta=\alpha+\beta + O(1/\eta)$. In this way, DRVM indeed tends to RVM within corrections synthetized by Eq.~(\ref{eq:DRVM}).
Therefore whatever the geometry of the off-centred dipole, at large distances, its observational signature is indiscernible from the centred dipole expectations. The only mean to disentangle between both models is by looking at emission from the vicinity of the stellar surface, like thermal X-ray emission for instance.

\subsection{Radio/X-ray time lag}

Thermal X-ray emission from the stellar surface helps to constrain the non-dipolar field components. In is paragraph, we derive the time lag between X-ray peak and radio peak in the off-centred dipole model.
The calculations performed in this paragraph help to understand the origin of the X-R time delay. We start with a toy model based purely on geometrical effects due to the shifted dipole. We end this paragraph with a discussion about the additional contribution from lensing and photon time of flight effects.

Radio emission becomes visible if the magnetic moment vector~$\bmu$ points towards the observer~$\nobs$. This condition translates into a time $t_{\rm n}$ such that $\gamma + \Omega\,t_{\rm n} = 0$ or more explicitly when 
\begin{equation}
\label{eq:tnorth}
 \frac{t_{\rm n}}{P} = -\frac{\gamma}{2\,\upi}
\end{equation}
corresponding to the visibility of the north pole. Symmetrically, the south pole becomes visible at a time $t_{\rm s}$ such that $\gamma + \Omega\,t_{\rm s} = \upi$ or more explicitly whenever 
\begin{equation}
\label{eq:tsouth}
\frac{t_{\rm s}}{P} = \frac{1}{2} - \frac{\gamma}{2\,\upi}.
\end{equation}
Thermal X-ray emission along the magnetic poles becomes visible with maximum intensity when the phase of the polar cap centre is located in the $xOz$ plane. This condition requires a phase $\phi_\pm=0$ meaning that the $y$-coordinate of the poles vanish whereas the $x$-coordinate $x_\pm>0$ (otherwise the pole would be hidden by the star), assuming that the observer line of sight lies in the $xOz$ plane. Let us call the $y$ coordinate of the north and south pole by~$y_+$ and $y_-$ respectively. Equations $\phi_\pm=0$ are solved analytically for the time lag between the peak in X-ray and radio for any geometry of the off-centred dipole. Explicitly the time-dependent $x$ and $y$ coordinates of both poles are given by
\begin{subequations}
\begin{align}
 x_\pm & = d \, \sin \delta \, \cos \Omega\,t + \lambda_\pm \, \sin \alpha \, \cos (\gamma+\Omega\,t) \\
 y_\pm & = d \, \sin \delta \, \sin \Omega\,t + \lambda_\pm \, \sin \alpha \, \sin (\gamma+\Omega\,t).
\end{align}
\end{subequations}
We are looking for the time $t_\pm$ satisfying $y_\pm(t_\pm)=0$. Because the dot product $\mathbf{m} \cdot \mathbf{d}$ is independent of time, the $y_\pm$ are a linear combination of two sinus functions with the same frequency and given by
\begin{subequations}
\label{eq:ypm}
\begin{align}
 x_\pm & = (A + B_\pm \,\cos\gamma ) \cos \Omega\,t - B_\pm \, \sin \gamma \sin \Omega\,t \\
 y_\pm & = B_\pm \, \sin \gamma \cos \Omega\,t + (A+B_\pm\,\cos\gamma) \sin \Omega\,t
\end{align}
\end{subequations}
where we introduced constants
\begin{subequations}
\begin{align}
A & = d \, \sin \delta \\
B_\pm & = \lambda_\pm \sin \alpha .
\end{align}
\end{subequations}
Expressions~(\ref{eq:ypm}) are recast into single trigonometric functions with standard techniques following the $\sin$ prescription. Therefore
\begin{subequations}
\begin{align}
 x_\pm & = R_\pm \, \cos(\Omega\,t-\psi_\pm) \\
 y_\pm & = R_\pm \, \sin(\Omega\,t-\psi_\pm)
\end{align}
\end{subequations}
where the new amplitudes~$R_\pm$ and phases~$\psi_\pm$ are given by
\begin{subequations}
\begin{align}
R_\pm & = \sqrt{A^2 + 2 \,A\,B_\pm\,\cos\gamma + B_\pm^2} \\
\tan\psi_\pm & = - \frac{B_\pm \, \sin \gamma}{A+B_\pm\,\cos\gamma} .
\end{align}
\end{subequations}
The $\tan$ leaves its argument~$\psi_\pm$ indefinite within an additional constant $k\,\upi$ with $k\in\mathbb{Z}$. This degeneracy is resolved by taking the angle in the proper quadrant, calling the $\arctan(x,y)$ function 
\begin{equation}
\label{eq:PsiArcTan}
\psi_\pm = \arctan(A+B_\pm\,\cos\gamma, -B_\pm \, \sin \gamma) .
\end{equation}
Some useful symmetries are recognized between both angles $\psi_+$ and $\psi_-$, namely
\begin{equation}
\psi_-(\upi-\alpha,\upi-\gamma,\delta,\epsilon) = - \psi_+(\alpha,\gamma,\delta,\epsilon)
\end{equation}
derived from the antisymmetry of
\begin{subequations}
\begin{align}
\lambda_-(\upi-\alpha,\upi-\gamma,\delta,\epsilon) & = - \lambda_+(\alpha,\gamma,\delta,\epsilon) \\
B_-(\upi-\alpha,\upi-\gamma,\delta,\epsilon) & = - B_+(\alpha,\gamma,\delta,\epsilon) .
\end{align}
\end{subequations}
The $y$ component of each hot spot vanishes if the normalised time is equal to
\begin{equation}
\frac{t_\pm}{P} = \frac{\psi_\pm}{2\,\upi} + \frac{k}{2}
\end{equation}
with $k\in\mathbb{Z}$. Moreover, the condition $x_\pm>0$ implies $k=0$ therefore
\begin{equation}
\frac{t_\pm}{P} = \frac{\psi_\pm}{2\,\upi} .
\end{equation}
The time lag between the radio pulse and the thermal X-ray light-curve maximum is therefore for each pole
\begin{subequations}
\begin{align}
\Delta_+ & = \frac{t_+-t_{\rm n}}{P} = \frac{\psi_+ + \gamma}{2\,\upi} \\
\Delta_- & = \frac{t_--t_{\rm s}}{P} = \frac{\psi_- + \gamma}{2\,\upi} - \frac{1}{2} .
\end{align}
\end{subequations}
The constant term $-1/2$ for the south pole arises because the observer will only see this pole half a period later compared to the north pole if they are perfectly antipodal. The time delay does not depend on the line of sight inclination~$\zeta$. The latter has only an impact on the light-curve shape and intensities but not on the longitude for which the flux is maximal.

In the limit of a small displacement from the centre of the star~$d\ll R$, the time delay reduces to first order in $\epsilon$ to
\begin{multline}
 \psi_+ = \arctan [ ( 1 - \epsilon \, \mathbf{m} \cdot \mathbf{d} ) \, \sin \alpha \cos \gamma + \epsilon \, \sin \delta , \\
 - ( 1 - \epsilon \, \mathbf{m} \cdot \mathbf{d} ) \, \sin \alpha \sin \gamma] .
\end{multline}
The time lag can also be computed from more geometrical considerations. Indeed, taking the angle between the projection of the magnetic moment onto the equatorial plane and the magnetic pole position vector leads to exactly the same result as before for the time lag between X-rays and radio.

Note that for the special case $\gamma=0\degr$, there is no time lag between both radio and X-ray light-curves, whatever the other parameters of the dipole. The same conclusion applies for the special case~$\delta=0\degr$. 

\begin{figure}
	\centering
	\includegraphics[width=0.5\textwidth]{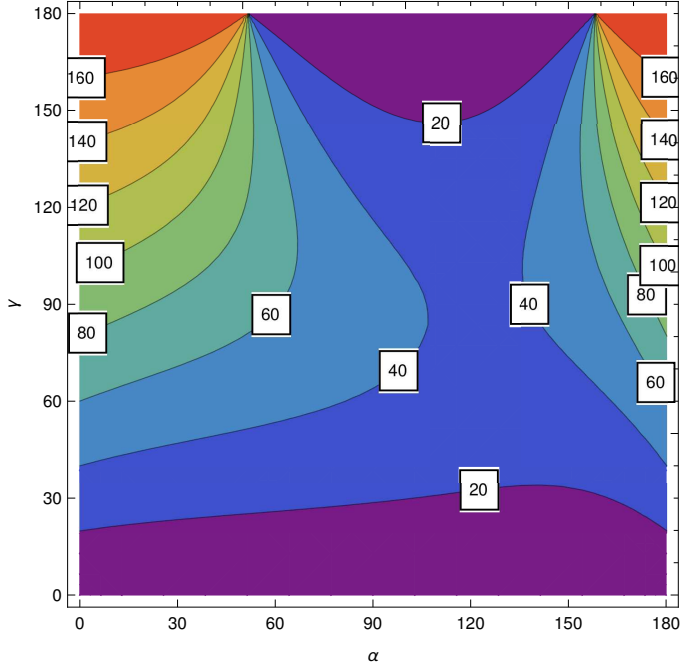}
	\caption{Time lag in degrees between X-ray and radio pulses for the north pole $\Delta_+$ depending on the angles~$\alpha$ and $\gamma$ for $\delta=110\degr$ and $\epsilon=0.8$.}
	\label{fig:RetardAlphaBeta}
\end{figure}

Fig.~\ref{fig:RetardAlphaBeta} shows a sample of time lags~$\Delta_+$ for the north pole depending on the angles~$\alpha$ and $\gamma$ of the off-centred dipole for $\delta=110\degr$ and $\epsilon=0.8$. These particular values are relevant for PSR~J1136+1551. The south pole time lags~$\Delta_-$ are founded by symmetry considerations.
We are able to reproduce time delay in the interval~$[-180\degr,180\degr]$ (negative values are obtained for $\gamma<0$ not shown in the plot) although half a period delay is only possible when $\alpha$ is nearly zero. Care must be taken for the special case of a nearly aligned rotator. A time lag of $P/2$ corresponding to $180\degr$ is possible but only for $\alpha\approx0\degr$. The time lag is maximal for an aligned or counter-aligned dipole ($\alpha\approx0\degr$ or $180\degr$). In these cases, the delay increase with the angle $\gamma$ to maximum for $\gamma=\pm180\degr$. The delay can be true retardation but also time advance if $\gamma<0$. For strongly inclined or almost orthogonal rotators, the maximal time lag is well below $P/2=180\degr$ and located around $\gamma=90\degr$.

\begin{figure}
	\centering
	\includegraphics[width=0.5\textwidth]{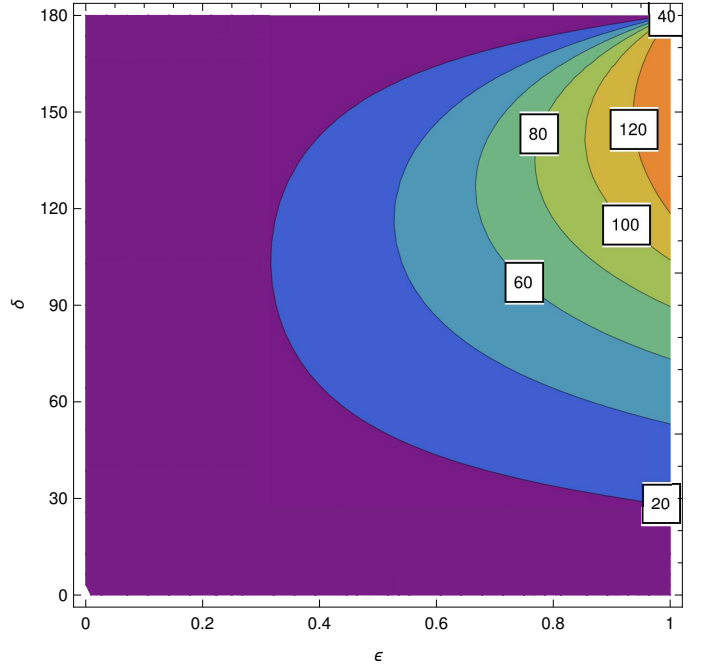}
	\caption{Time lag in degrees for the north pole $\Delta_+$ for varying~$\delta$ and $\epsilon$ for $\alpha=130\degr$ and $\zeta=134.2\degr$.}
	\label{fig:RetardDeltaEpsilon}
\end{figure}

In Fig.~\ref{fig:RetardDeltaEpsilon} time lags~$\Delta_+$ for the north pole are shown depending on the angle~$\delta$ and on the normalized displacement~$\epsilon$ for $\alpha=130\degr$ and $\zeta=134.2\degr$. For almost centred dipole, the lag is negligible as expected and increases when the dipole is shifted closer and closer to the surface for a given $\delta$. Again, the south pole delay~$\Delta_-$ is founded by symmetry considerations.

The above estimates rely only on geometrical effects without light bending or Shapiro delay or retardation. Let us now quantify these contributions with respect to the previous estimate. For lensing, we employ the Schwarzschild light bending formula relating the impact parameter~$b$
\begin{equation}
b = \dfrac{r}{\sqrt{1-\frac{\Rs}{r}}} \, \sin A
\label{eq:impact}
\end{equation}
to the variation in angle~$\Delta \chi$ by integration of \citep{pechenick_hot_1983}
\begin{equation}
\label{eq:bending}
\Delta \chi(r) = \pm \int_{r_{0}}^{r} \dfrac{b\, dr}{r^{2}\sqrt{1-\frac{b^{2}}{r^{2}}(1-\frac{\Rs}{r})}}
\end{equation}
where $A$ represents the angle between the photon direction at emission site at a distance~$r$ and the radial direction. The Shapiro time delay induced by this curved path is
\begin{equation}
\label{eq:shapiro_exact}
c \, \Delta t(r) =  \pm \int_{r_0}^r \frac{dr}{\left( 1 - \frac{R_{\rm s}}{r} \right) \, \sqrt{1 - \left( 1 - \frac{R_{\rm s}}{r} \right) \, \frac{b^2}{r^2} }}
\end{equation}
the sign in front of the integrals depending ton the receding or approaching photon trajectory.

\begin{figure}
	\centering
	\includegraphics[width=0.9\linewidth]{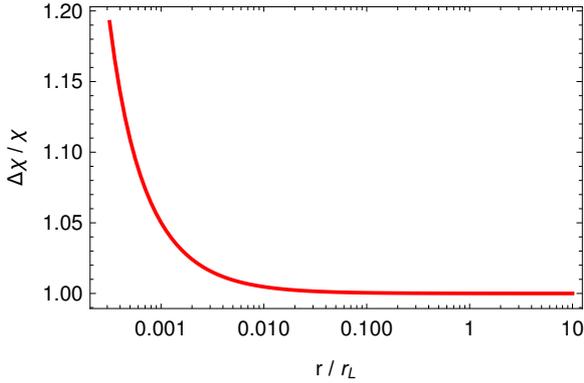}
	\caption{Light bending ratio $\Delta\chi/\chi$ obtained by integration of Eq.~(\ref{eq:bending}) for PSR~J1136+1551 for different emission heights $r$ and an observer placed at a distance $D=10^6 \, \rlight$.}
	\label{fig:bending}
\end{figure}

\begin{figure}
\centering
\includegraphics[width=0.9\linewidth]{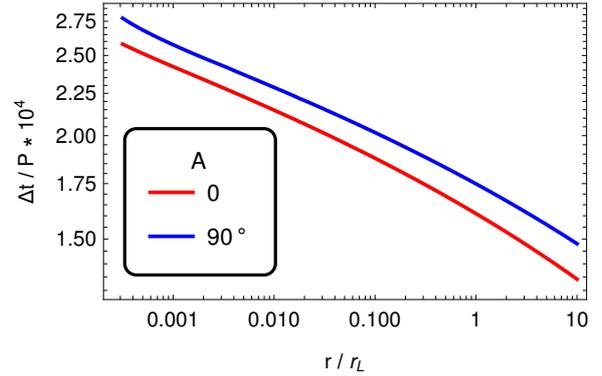}
\caption{Extra time delay induced by Shapiro delay $\Delta t/P$, obtained by integration of Eq.~(\ref{eq:shapiro_exact}) and normalized to the period of PSR~J1136+1551 for an observer placed at a distance $D=10^6 \, \rlight$. Note the factor $10^4$ in the normalisation.}
\label{fig:shapiro_norme}
\end{figure}

Thermal X-rays emanate from the polar caps as an isotropic emission, with maximum flux perpendicular to the stellar surface, thus in the radial direction with $A=0$ and $r=R$. We therefore do not expect any light bending ($\chi=0$) for the rays at maximum intensity. This is an exact result relying on eq.~(\ref{eq:bending}). However, the Shapiro time delay for a straight motion to a distance $D$ is given by
\begin{equation}\label{eq:Shapiro_Thermal}
 c\, \Delta t = D-R + \Rs \, \ln\left(\frac{D-\Rs}{R-\Rs}\right)
\end{equation}
the log term showing the influence of gravity. Moreover, as will be shown for PSR~J1136+1551, radio emission is produced at high altitude, well above the polar caps for which $r_{\rm radio}\gg R$. The ray is not directed into the radial direction due to the off-centring. In such a case, the strongest Shapiro delay arises for an angle~$A=90\degr$. The impact parameter then reduced to the minimal approach distance $r_{\rm radio}$. First order corrections in $\Rs$ then give
\begin{equation}
\label{eq:Shapiro_minimal_approach}
\begin{aligned}
c \, \Delta t & \approx \sqrt{D^2-r_{\rm radio}^2} + \Rs \, \ln\left( \frac{D + \sqrt{D^2-r_{\rm radio}^2}}{r_{\rm radio}} \right) \\
 & + \frac{\Rs}{2} \, \sqrt{\frac{D-r_{\rm radio}}{D+r_{\rm radio}}} .
\end{aligned}
\end{equation}
The first term on the right hand side $\sqrt{D^2-r_{\rm radio}^2}$ corresponds to flat spacetime propagation.
Light bending of radio photons at the emission height of several tenths of stellar radii is negligible, even for a maximum angle of $A=90\degr$. The plot in Fig.~\ref{fig:bending} showing the ratio $\Delta\chi/\chi$ clearly demonstrates that for $r/\rlight \gtrsim 0.01$ corrections are small, photons are almost not deflected. The observer is located at a distance $D=10^6 \, \rlight$. We can safely use flat spacetime retardation effects.
The extra time added by Shapiro delay is shown in Fig.~\ref{fig:shapiro_norme} for parameters relevant to PSR~J1136+1551. We considered to extreme cases: a straight line propagation with $A=0\degr$ and a maximally bent trajectory with $A=90\degr$. In general, for normal radio pulsars with period $P\gtrsim100$~ms, the spacetime curvature delay is irrelevant, amounting to a tiny fraction of $10^{-4}$ of the period~$P$.

From all the above study, it appears that general-relativistic effects can be discarded when photon propagation is concerned. Simple flat spacetime estimates are sufficient to very good accuracy for slowly rotating neutron stars with $P\gtrsim100$~ms. Consequently, besides geometrical effects explained in detail at the beginning of this paragraph, an additional delay must be taken into account via the time of flight between the thermal emission site and the radio emission site. Normalized to the period of the star, we get \citep{petri_unified_2011}
\begin{equation}\label{eq:time_of_flight}
 \frac{\Delta t}{P} = \frac{R-r_{\rm radio}}{2 \,\upi\,\rlight} \lessapprox 0.03 .
\end{equation}
So again, the propagation effect can only account for a few \% of the X-R time delay, largely below the time delay measured in PSR~J1136+1551.

\subsection{Hot spot light-curves}

The above calculations do not take into account general-relativistic effects like Shapiro delay and light bending. However, the neutron star compactness defined by the ratio between Schwarzschild radius~$\Rs$ and stellar radius~$R$, computed by~$K=\Rs/R$ is far from negligible and about $K\approx0.41$ for standard parameters of size $R=10$~km and mass $M=1.4~M_\odot$. Accurate computations of these effects would require path integrations in Schwarzschild or Kerr metric but for a rapid estimate on the off-centred hot spot light-curves, we use the approximation found by \cite{beloborodov_gravitational_2002} and summarized by the observed flux from the north pole
\begin{equation}
\label{eq:Flux}
 f_{\rm n} = 
\begin{cases}
\left( 1 - K \right) \, \cos i + K & \text{ if } \cos i > - \frac{K}{1-K} \\
0 & \text{ if } \cos i < - \frac{K}{1-K}
\end{cases}
\end{equation}
and from the south pole
\begin{equation}
 f_{\rm s} = 
\begin{cases}
-\left( 1 - K \right) \, \cos i + K & \text{ if } \cos i < \frac{K}{1-K} \\
0 & \text{ if } \cos i > \frac{K}{1-K}
\end{cases}
\end{equation}
The angle~$i$ represents the angle between the normal to the hot spot~$\mathbf{n}_{\rm pc}$ and the line of sight and is therefore given by
\begin{equation}
 \label{eq:cosi}
 \cos i = \mathbf{n}_{\rm pc} \cdot \nobs .
\end{equation}
Note that these expressions hold only for a centred dipole when both poles are symmetrically located with respect to the stellar centre.

The two hot spots become visible if the angle between the normal to the north hot spot surface and the line of sight becomes less than
\begin{equation}
\cos i = \frac{K}{1-K} .
\end{equation}

\begin{figure}
	\centering
	\includegraphics[width=0.5\textwidth]{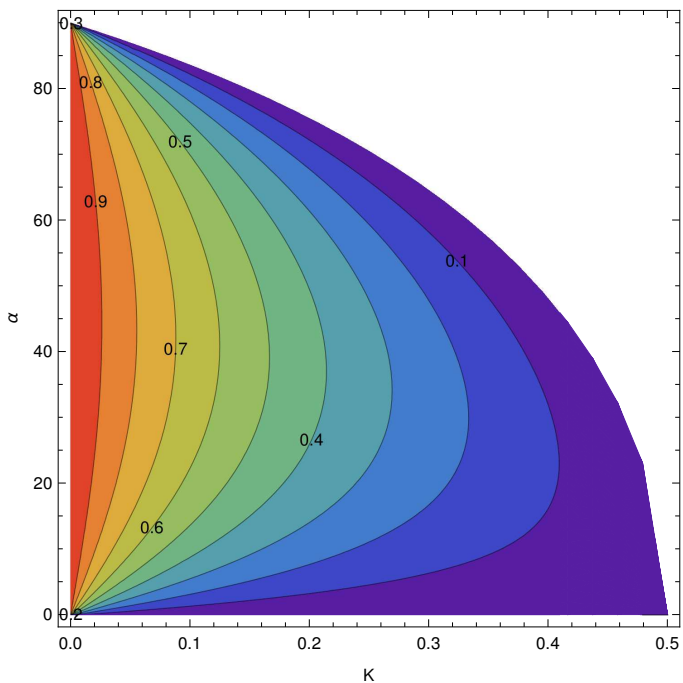}
	\caption{Maximum pulsed fraction depending on obliquity~$\alpha$ and compactness~$K$.}
	\label{fig:DeuxPointChaudImpossible}
\end{figure}

Fig.~\ref{fig:DeuxPointChaudImpossible} shows the maximum pulsed fraction depending on obliquity~$\alpha$ and compactness~$K$. It demonstrates the impossibility to see only one hot spot with a significant pulsed fraction when the spots are antipodal and with realistic compactnesses of $K\gtrsim0.3$. With such compactness, the pulsed fraction is at most 15\%. 

According to the X-ray light curves of PSR~J1136+1551, there are strong hints that the two hot spots are neither antipodal nor symmetric. In the next section, we show how to constrain the geometry of the hot spots of PSR~J1136+1551 to agree with the radio polarization angle profile simultaneously with the X-ray light-curves delayed by about $60\degr$ with respect to the radio pulse profile.

In the work of \cite{annala_constraining_2010}, more than 100 X-ray pulse profiles were analysed to constrain their compactness and geometry. They found that for a centred dipole, 79\% should be double peaked, implying an obliquity of $\alpha<40\degr$. This strongly suggests that the hot spots are neither identical nor antipodal as often claimed.

\section{Case study: PSR~J1136+1551}
\label{sec:psr_j1136}

PSR~J1136+1551 is the perfect target for our study. It is a slowly rotating pulsar with excellent radio polarization data and fairly good X-ray spectra and light-curves. Table~\ref{tab:PSRJ1136} summarizes its main observed properties. With a period of $P=1.19$~s, its polar caps are much smaller that the radio pulse profile width. We have to extend the emissivity directivity that is go to higher altitudes because magnetic field lines diverge. Simple geometric arguments lead to an altitude of several hundred of kilometres. Indeed the pulse width, denoted by~$W$ is about 4\% of the period or expressed in radians $W = 0.04 \times \,2\upi$. But, assuming an aligned dipole, the opening angle is related to the position by  $W = 3\, \theta_{\rm W}/2$. The radial distance is therefore $r = \rlight\,\sin^2\theta_{\rm W} \approx 1576$~km. 
In fact more rigorous methods applied to estimate radio emission heights as shown in section~\ref{sec:radio_obs} limits the emission to
originate at slightly lower heights of around 400~km,  
which is still well within the light-cylinder but at sufficiently high altitude to minder the effect of an off-centring.
\begin{table}
	\centering
	\begin{tabular}{lc}
		\hline
		Period (s) & $1.187913065936$ \\
		Period derivative (s/s) & $3.733837\times10^{-15}$ \\
		Distance (pc) & 357 \\
		Obliquity~$\alpha$ & 50\degr/130\degr \\
		Line of sight~$\zeta$ & 54\degr/134\degr \\
		BB temperature &  $2.9^{+0.6}_{-0.4}$~MK \\ 
		BB fraction & 0.45 \\
		BB luminosity & $2.4\times10^{28}$~ergs s$^{-1}$ \\
		Polar cap radius & $14^{+7}_{-5}$~m \\
		\hline
	\end{tabular}
	\caption{\label{tab:PSRJ1136}Main observed and inferred characteristics of PSR~J1136+1551 with the two possible orientations for $\alpha$ and $\zeta \approx \alpha + 4.2 \degr$.}
\end{table}

S17 finds that the X-ray spectrum can be fitted with a black body and a power law. The black body dominates in the energy range 0.5 to 1.2~keV and can be fitted with temperatures of $2.9^{+0.6}_{-0.4}$~MK and radius of $14^{+7}_{-5}$~m, corresponding to black body luminosity of about $2.4\times10^{28}$~ergs s$^{-1}$. The fraction of the black body in the best fit (using Table~4 of S17) spectrum is about 0.45.
The distance is taken from \cite{brisken_very_2002}. 
As shown by S17 the fitted black body area and temperature is consistent with the partially screened inner vacuum gap model (PSG model, see \cite{gil_drifting_2003}). One essential interpretation of the smaller than dipolar polar cap area obtained from black body fit is the presence of strong multipolar surface magnetic fields. If one accepts this interpretation, then it is expected that the polar cap is located at a different location compared to the star centred dipole axis. This motivates us to consider the offset dipole model as a first order approximation for the multipolar field.

\subsection{Thermal emission}

PSR~J1136+1551 requires two hot spots that are not antipodal from which we compute the approximate flux. Such geometry is easily derived from an off-centred dipole. We therefore straightforwardly extend \cite{beloborodov_gravitational_2002} work to any hot spot geometry as follows.

Define the two hot spots with their spherical coordinates such that the north pole is at ($\theta_{\rm n},\phi_{\rm n}$) and the south pole at ($\theta_{\rm s},\phi_{\rm s}$). These positions define the unit vectors $\mathbf{n}_{\rm n}$ and $\mathbf{n}_{\rm s}$ along the north and south pole respectively. From X-ray observations, the south pole should never be seen because of the sinusoidal shape of the light-curve or less stringently much weaker than the north spot. This puts some constrain on $\theta_{\rm s}$ 
because coming back to the definition of the angle $\cos i$ in Eq.(\ref{eq:cosi}), the south pole remain invisible whenever
\begin{equation}
\cos i_{\rm s} = \mathbf{n}_{\rm s} \cdot \nobs < \frac{K}{K-1} \approx -\frac{1}{2}
\end{equation}
where we assumed $K\approx1/3$ for the last number. Thus the angle $i_{\rm s}$ must be larger than $120\degr$. But this angle $\cos i_{\rm s}$ remains between $\cos(\theta_{\rm s}+\zeta)$ and $\cos(\theta_{\rm s}-\zeta)$. From geometrical considerations, we get additional constraints such that
\begin{subequations}
\begin{align}
	\theta_{\rm s} & > \zeta + \arccos \frac{K}{K-1} \\
	\zeta & < \upi - \arccos \frac{K}{K-1}
\end{align}
\end{subequations}
These constraints are not easily satisfied if the two hot spots were antipodal and symmetric. We pin down the geometry of PSR~J1136+1551 by a combined radio and X-ray fitting as explained in the following lines.

Note that the X-ray light-curves computed from Eq.~(\ref{eq:Flux}) do only depend on $\cos i$ found from Eq.~(\ref{eq:cosi}). The normal to the polar caps are directed along $\mathbf{r}_\pm$ given in Eq.~(\ref{eq:pole}). However, the configuration is degenerate in the sense that that any new position~$\mathbf{d}'$ of the magnetic moment, deduced from $\mathbf{d}$ by
\begin{equation}\label{key}
 \mathbf{d}' = \mathbf{d} + a \, \lambda_\pm \, \mathbf{m}
\end{equation}
with $a<1$ would give the same light-curves. Consequently, there is a freedom in choosing the location of the magnetic moment along the direction pointed by $\mathbf{m}$. This indeterminacy can only be removed if microphysics is included (but out of the scope of this work based on pure geometrical considerations). Physically, this means that the magnetic moment can be brought closer to one or another hot spot and influence the luminosity. We will come back to this later.

\begin{figure}
	\centering
	\resizebox{\columnwidth}{!}{\input{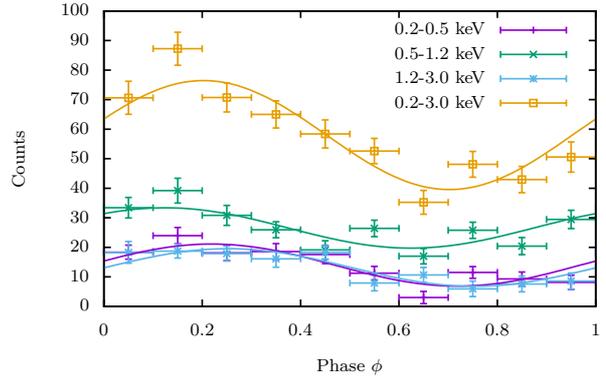}}
	\caption{X-ray light-curves with error bars (crosses) and best fitting parameters (solid lines) for an obliquity~$\alpha = 50\degr$ or $\alpha = 130\degr$ and a compactness~$K=0.35$.}
	\label{fig:Xray_light_curves}
\end{figure}

From the radio polarization data, we known that the two possible orientations are $\alpha=50\degr$ or $\alpha=130\degr$ with $\zeta=\alpha+4.2\degr$. We use these constrains to fit independently the X-ray light-curves shown in Fig.\ref{fig:Xray_light_curves} for several energy bands: 0.2-0.5~keV, 0.5-1.2~keV, 1.2-3.0~keV and the full band 0.2-3.0~keV. The expression for the flux is given by Eq.~(\ref{eq:Flux}) for one spot, disregarding the second spot. We need to find the amplitude~$A$ of the flux, the longitude shift~$\gamma$ and the location of the dipole depicted by $\delta$ and $\epsilon$ independently in each energy band. The best parameters found by a $\chi^2$ adjustment are summarized in Table~\ref{tab:Fit} separately for the individual bands and the total flux. For both orientations with $\alpha=50\degr$ or $\alpha=130\degr$, the offset is very similar, close to the stellar surface at about $\epsilon \approx 0.7-0.9$ except for the band $0.5-1.2$~keV requiring a lower offset, with a shift in longitude $\gamma \approx 120\degr - 130\degr$ but with different positions for the magnetic moment, around $\delta\approx 45\degr-65\degr$ for $\alpha=50\degr$ but around $\delta\approx 105\degr-125\degr$ for $\alpha=130\degr$ thus about the complementary angle $180\degr-\delta$ for the second geometry. The two orientations show the most likely parameters to fit X-R and X-ray light-curves. Indeed the fits are equally good irrespective of the energy band considered.

\begin{table}
\begin{tabular}{cccccc}
\hline
$\alpha$ & Band (keV) & A & $\gamma$ & $\delta$ & $\epsilon$ \\
\hline
$50\degr$ 
& 0.2-0.5 & 21. & 113. & 65. & 0.83 \\
& 0.5-1.2 & 36.	& 140. & 46. & 0.59 \\
& 1.2-3.0 & 20.	& 118. & 60. & 0.89 \\
& 0.2-3.0 & 81.	& 129. & 53. & 0.74 \\
\hline
$130\degr$ 
& 0.2-0.5 & 21.	& 106.	& 108.	& 0.86 \\
& 0.5-1.2 & 34.	& 132.	& 128.	& 0.57 \\
& 1.2-3.0 & 19.	& 112.	& 114.	& 0.92 \\
& 0.2-3.0 & 77.	& 122.	& 122.	& 0.75 \\
\hline
\end{tabular}
\caption{Best fit parameters $(A, \gamma, \delta, \epsilon)$ for X-ray light-curves with $\alpha = 50\degr$ or $\alpha = 130\degr$ for the different energy bands.}
\label{tab:Fit}
\end{table}

\subsection{Polar cap geometry}

\begin{figure}
	\centering
	\resizebox{\columnwidth}{!}{\input{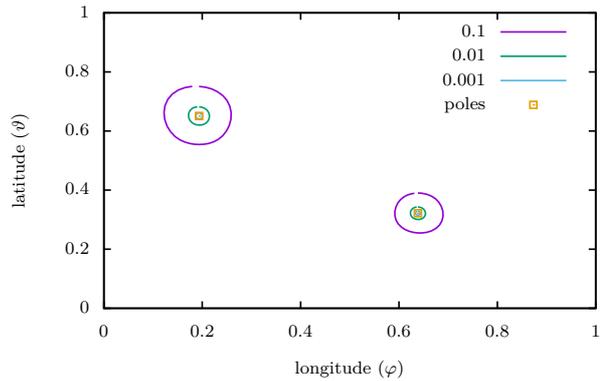}}
	\caption{Map of the polar cap shapes and magnetic pole location on the surface of the star for different spin rates $R/\rlight=\{0.001,0.01,0.1\}$ with $(\alpha, \gamma, \delta, \epsilon) = (60\degr,60\degr,60\degr,0.3)$.} 
	\label{fig:PolarCapShapeAndPoles}
\end{figure}

What happens to the second hot spot? In the configurations found above, it should also be visible. However, the off-centred dipole has a strong impact on the polar cap shape. In Fig.~\ref{fig:PolarCapShapeAndPoles} we show the rim of the polar caps and the location of the magnetic poles for $(\alpha, \gamma, \delta, \epsilon)=(60\degr,60\degr,60\degr,0.3)$ and different spin rates with $R/\rlight=\{0.001,0.01,0.1\}$. These are the geometric localisation of the last closed magnetic field foot points on the surface. The size of the polar cap scales approximately as $\sqrt{R/\rlight}$ as for an aligned rotator.

\begin{figure}
	\centering
	\resizebox{\columnwidth}{!}{\input{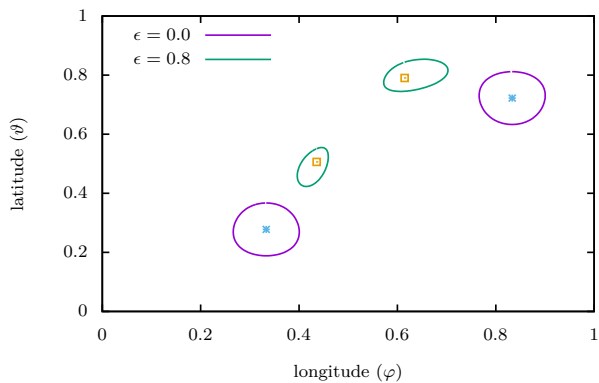}}
	\caption{Polar cap shape and magnetic pole location depending on aberration and retardation for centred and off-centred dipole rotating at a spin rate $R/\rlight=0.1$.}
	\label{fig:PolarCapShapeAberration}
\end{figure}

If the second configuration with $\alpha=130\degr$ is kept, the two polar cap rims are very different as shown in Fig.~\ref{fig:PolarCapShapeAberration}. They possess a very different size, the second being much smaller (note the size must be scaled down to $\sqrt{R/\rlight}$ for PSR~J1136+1551 however the ratio remains the same). Therefore, the second hot spot is much fainter than the primary hot spot because of its smaller area but probably also because of its lower electromagnetic activity and polar cap heating implied by its smaller size. We expect therefore the second spot to be much fainter and drowned in the larger hot spot signal. 

There are several reasons to expect asymmetrical emission properties from both polar caps. The first one is that one hot spot is several times smaller than the other hot spot because of the geometry of the off-centred dipole. The second one is related to the relative distance $D_\pm$ of the magnetic moment~$\bmu$ with respect to the stellar surface where the poles are located. In the general case, one hot spot is closer to the magnetic moment than the other spot. In such a case, because of the $D_\pm^{-3}$ decrease of the magnetic dipolar field strength, its intensity at both polar caps can be very different, scaling like $(D_+/D_-)^3$ where $D_\pm$ are the distances of the magnetic moment to each hot spot. This implies a larger curvature, therefore larger accelerating electric fields and higher magnetic photo-absorption and therefore more numerous and more energetic particles for the hot spot closest to the magnetic moment. Consequently, this hot spot will appear much brighter than the other hot spot.

Consequently, both hot spots being visible does not contradict the fact that only the most brilliant is detected. The strong asymmetry in polar cap shape and size spoils any attempt to fit solely thermal X-rays from hot spots in hope to constrain neutron star mass over radius ratio. A multi-wavelength approach is much more fruitful as demonstrated in this paper.


\subsection{The relevance of an off-centred dipole}

The above study showed that the radio and X-ray light-curves and polarization properties are best fitted with an external off-centred dipole located very close to the surface of the star, only a few kilometres or less. This shift should not be misinterpreted as a real dipole existing inside the star. It is well known that a centred magnetic dipole filling vacuum outside a perfect spherical conductor can also be produced by an internal uniform and homogeneous magnetization. In the same vein, an off-centred dipole in vacuum can be accounted for with a heterogeneous magnetization inside the star showing a strong spatial gradient. Moreover, the core of a neutron star being certainly superconductor, the magnetic field is expelled to the outer edge, anchored in the crust, drastically modifying the dipolar configuration inside. Our shifted dipole inside the star is only intended to generate a simple non-dipolar component with the least number of free parameters. There is no physical reason to keep a dipole inside the star.

Moreover, the origin of neutron star magnetic fields is not accurately known but it is believed to be produced partly by the magnetic flux freezing during the core collapse of the progenitor \citep{woltjer_x-rays_1964} and/or by the combination of convection and differential rotation inside the star \citep{thompson_neutron_1993}. Crustal thermomagnetic effects have also been invoked \citep{blandford_thermal_1983} \citep{urpin_generation_1986}. It is known that a purely poloidal or toroidal magnetic field is unstable \citep{markey_adiabatic_1974} \citep{flowers_evolution_1977} and that a combined poloidal/toroidal configuration is required \citep{wright_pinch_1973}. But the details of the interaction are not well known to date. Lastly, the evolution of an off-centred magnetic field is not expected to show large discrepancies with respect to a centred dipole as its decay or increase is mostly related to the physics of the crust in which it is anchored and on the accreting matter like a fall-back disk if any. However the time scale and mechanisms responsible for this decay or increase are still debated.

\section{Non-Thermal Emission}
\label{sec:non_thermal_emission}

X-ray spectra cannot conclusively distinguish between thermal and non-thermal emission, and thus
we need to consider what the X-R offset would mean in case the X-ray emission is non-thermal in nature.
Whether X-ray photons are produced by a thermal or a non thermal mechanism strongly affects the expected location of their emission sites. Almost black body radiation is well constrained to emanate from the polar cap thus at zero altitude from the surface. In this first case, the relative position between radio and X-ray production sites are well known. However, if the X-ray spectrum shows a non-thermal component, the picture becomes less clear. In this second case, photons must be produced within the magnetosphere or even within the wind, at a significant altitude above the neutron star surface, a significant fraction of $\rlight$. The time lag between radio and X-ray pulse profiles then strongly depends on the relative altitude between both emission sites. If non-thermal X-ray photons are coming from regions above the radio emission height, and directed along open field lines, we would perceive X-rays before radio photons. On the contrary, if these non-thermal X-rays are coming from regions below the radio emission height, we would perceive X-rays after radio photons. This is simply due to propagation effects like time of flight. As shown by \cite{petri_unified_2011}, this time lag~$\Delta t$ corresponds to a fraction of the pulsar period given by
\begin{equation}\label{eq:TOF}
 \frac{\Delta t}{P} = \frac{\Delta h}{2\,\upi\,\rlight}
\end{equation}
where $\Delta h = h_{\rm radio} - h_{\rm X}$ denotes the difference in altitude between radio height~$h_{\rm radio}$ and X-ray height~$h_{\rm X}$. It can be positive or negative, accounting for a delay or advance in time of X-ray reception with respect to the radio signal reception. For PSR~J1136+1551, we know that $h_{\rm radio}/\rlight \approx 0.027$, thus the time lag must be smaller than $\Delta t/P \lesssim 1/(2\,\upi) \approx 0.16$ corresponding to a maximum phase shift of~$57\degr$. However, the phase shift measured in PSR~J1136+1551 is close to or slightly above this value. We conclude that a non-thermal origin of the X-ray is highly disfavoured to explain the X-R shift.


\section{Conclusions}
\label{sec:conclusion}

We showed that a combined radio and X-ray light-curve fitting is a powerful tool to disentangle the degeneracy between several geometric configurations. Indeed, radio observation and polarization is able to precisely locate the radio emission altitude but not the geometry of viewing angle and obliquity. Non is the X-ray data alone able to put severe constrain on this geometry. However, their simultaneous modelling allows to pin down this geometry to good accuracy. We showed by an example of PSR~J1136+1551 that the time lag between X-ray and radio is naturally explained by an off-centred dipole locate close to the surface of the star. In reality however the exact nature of the magnetic field can be more complex, and to model such complex magnetic field structure is difficult and various other observational constrains need to be invoked, which is beyond the scope of this work. The offset dipole model for the magnetic field considered in this work, is the simplest approximation of non dipolar magnetic field which clearly demonstrates that non dipolar magnetic fields are probably ubiquitous on the surface. 

In high altitude emission sites such that $h\gg R$, the difference between off-centred and centred is smeared out and in principle the same fit applies to the DRVM. It is impossible to constrain the DRVM when radio photons are produced or leaves the system at large distance. Only the millisecond pulsars are able to disentangle between RVM and DRVM when $h\lesssim R$ but in such cases, the aberration/retardation effect is no more valid and the non dipolar fields already enter the game in the radio emission. We are therefore at a too early stage to fit millisecond pulsars.

In the future, we plan to investigate other slowly rotating pulsars seen in radio and X-ray to fit their geometry. If also seen in gamma-ray, it will help to localise the production sites of high energy photons in MeV/GeV range within the light-cylinder or within the wind.

\section*{Acknowledgements}

We are grateful to the referee for helpful comments and suggestions. We would like to thank Dr. Andrzej Szary for providing us with the data used in Fig. 12. We thank the staff of the GMRT who have made these observations possible. The GMRT is run by the National Centre for Radio Astrophysics of the Tata Institute of Fundamental Research. DM would like to thank Universit\'e de Strasbourg, CNRS, Observatoire astronomique de Strasbourg for hosting his visit where a large portion of this work was completed. This work has been supported by CEFIPRA grant IFC/F5904-B/2018.
J. P\'etri would like to acknowledge the High Performance Computing
center of the University of Strasbourg for supporting this work by
providing scientific support and access to computing resources. Part of
the computing resources were funded by the Equipex Equip@Meso project
(Programme Investissements d'Avenir) and the CPER Alsacalcul/Big Data.

\bsp	
\label{lastpage}

\begin{thebibliography}{39}
	\expandafter\ifx\csname natexlab\endcsname\relax\def\natexlab#1{#1}\fi
	
	\bibitem[{Annala \& Poutanen(2010)}]{annala_constraining_2010}
	Annala M., Poutanen J., 2010, Astronomy and Astrophysics, 520, A76
	
	\bibitem[{Arumugasamy \& Mitra(2019)}]{arumugasamy_evaluating_2019}
	Arumugasamy P., Mitra D., 2019, Monthly Notices of the Royal Astronomical
	Society, 489, 4589
	
	\bibitem[{Beloborodov(2002)}]{beloborodov_gravitational_2002}
	Beloborodov A.~M., 2002, ApJ, 566, L85
	
	\bibitem[{Blandford {et~al}\mbox{.}(1983)Blandford, Applegate, \&
		Hernquist}]{blandford_thermal_1983}
	Blandford R.~D., Applegate J.~H., Hernquist L., 1983, Mon Not R Astron Soc,
	204, 1025
	
	\bibitem[{{Blaskiewicz} {et~al}\mbox{.}(1991){Blaskiewicz}, {Cordes}, \&
		{Wasserman}}]{1991ApJ...370..643B}
	{Blaskiewicz} M., {Cordes} J.~M., {Wasserman} I., 1991, \apj, 370, 643
	
	\bibitem[{Brisken {et~al}\mbox{.}(2002)Brisken, Benson, Goss, \&
		Thorsett}]{brisken_very_2002}
	Brisken W.~F., Benson J.~M., Goss W.~M., Thorsett S.~E., 2002, ApJ, 571, 906
	
	\bibitem[{Deutsch(1955)}]{deutsch_electromagnetic_1955}
	Deutsch A.~J., 1955, Annales d'Astrophysique, 18, 1
	
	\bibitem[{{Dyks}(2008)}]{2008MNRAS.391..859D}
	{Dyks} J., 2008, \mnras, 391, 859
	
	\bibitem[{{Dyks} \& {Harding}(2004)}]{2004ApJ...614..869D}
	{Dyks} J., {Harding} A.~K., 2004, \apj, 614, 869
	
	\bibitem[{Dyks \& Harding(2004)}]{dyks_rotational_2004}
	Dyks J., Harding A.~K., 2004, The Astrophysical Journal, 614, 869
	
	\bibitem[{{Everett} \& {Weisberg}(2001)}]{2001ApJ...553..341E}
	{Everett} J.~E., {Weisberg} J.~M., 2001, \apj, 553, 341
	
	\bibitem[{Flowers \& Ruderman(1977)}]{flowers_evolution_1977}
	Flowers E., Ruderman M.~A., 1977, The Astrophysical Journal, 215, 302
	
	\bibitem[{Gil {et~al}\mbox{.}(2003)Gil, Melikidze, \&
		Geppert}]{gil_drifting_2003}
	Gil J., Melikidze G.~I., Geppert U., 2003, {\textbackslash}aap, 407, 315
	
	\bibitem[{{Gil} {et~al}\mbox{.}(2002){Gil}, {Melikidze}, \&
		{Mitra}}]{2002A&A...388..235G}
	{Gil} J.~A., {Melikidze} G.~I., {Mitra} D., 2002, \aap, 388, 235
	
	\bibitem[{{Hibschman} \& {Arons}(2001)}]{2001ApJ...546..382H}
	{Hibschman} J.~A., {Arons} J., 2001, \apj, 546, 382
	
	\bibitem[{Kargaltsev {et~al}\mbox{.}(2006)Kargaltsev, Pavlov, \&
		Garmire}]{kargaltsev_x-ray_2006-1}
	Kargaltsev O., Pavlov G.~G., Garmire G.~P., 2006, ApJ, 636, 406
	
	\bibitem[{Lyutikov(2016)}]{lyutikov_relativistic_2016}
	Lyutikov M., 2016, arXiv:1607.00777 [astro-ph], arXiv: 1607.00777
	
	\bibitem[{Markey \& Tayler(1974)}]{markey_adiabatic_1974}
	Markey P., Tayler R.~J., 1974, Mon Not R Astron Soc, 168, 505
	
	\bibitem[{Mitra {et~al}\mbox{.}(2016)Mitra, Basu, Maciesiak, Skrzypczak,
		Melikidze, {Andrzej Szary}, \& Krzeszowski}]{mitra_meterwavelength_2016}
	Mitra D., Basu R., Maciesiak K., Skrzypczak A., Melikidze G.~I., {Andrzej
		Szary}, Krzeszowski K., 2016, ApJ, 833, 28
	
	\bibitem[{{Mitra} \& {Li}(2004)}]{2004A&A...421..215M}
	{Mitra} D., {Li} X.~H., 2004, \aap, 421, 215
	
	\bibitem[{{Mitra} \& {Rankin}(2002)}]{2002ApJ...577..322M}
	{Mitra} D., {Rankin} J.~M., 2002, \apj, 577, 322
	
	\bibitem[{Pechenick {et~al}\mbox{.}(1983)Pechenick, Ftaclas, \&
		Cohen}]{pechenick_hot_1983}
	Pechenick K.~R., Ftaclas C., Cohen J.~M., 1983, The Astrophysical Journal, 274,
	846
	
	\bibitem[{{P{\'e}tri}(2017)}]{2017MNRAS.466L..73P}
	{P{\'e}tri} J., 2017, \mnras, 466, L73
	
	\bibitem[{Pétri(2011)}]{petri_unified_2011}
	Pétri J., 2011, MNRAS, 412, 1870
	
	\bibitem[{Pétri(2012)}]{petri_pulsar_2012}
	Pétri J., 2012, MNRAS, 424, 605
	
	\bibitem[{Pétri(2016)}]{petri_radiation_2016}
	Pétri J., 2016, MNRAS, 463, 1240
	
	\bibitem[{Pétri(2017)}]{petri_polarized_2017}
	Pétri J., 2017, MNRAS, 466, L73
	
	\bibitem[{Radhakrishnan \& Cooke(1969)}]{radhakrishnan_magnetic_1969}
	Radhakrishnan V., Cooke D.~J., 1969, Ap. Lett., 3, 225
	
	\bibitem[{{Rankin}(1983)}]{1983ApJ...274..333R}
	{Rankin} J.~M., 1983, \apj, 274, 333
	
	\bibitem[{{Rankin}(1993)}]{1993ApJ...405..285R}
	{Rankin} J.~M., 1993, \apj, 405, 285
	
	\bibitem[{Shitov(1983)}]{shitov_period_1983}
	Shitov Y.~P., 1983, Soviet Astronomy, 27, 314
	
	\bibitem[{Spitkovsky(2006)}]{spitkovsky_time-dependent_2006}
	Spitkovsky A., 2006, ApJ, 648, L51
	
	\bibitem[{Szary {et~al}\mbox{.}(2017)Szary, Gil, Zhang, Haberl, Melikidze,
		Geppert, {Dipanjan Mitra}, \& Xu}]{szary_xmm-newton_2017}
	Szary A., Gil J., Zhang B., Haberl F., Melikidze G.~I., Geppert U., {Dipanjan
		Mitra}, Xu R.-X., 2017, ApJ, 835, 178
	
	\bibitem[{Thompson \& Duncan(1993)}]{thompson_neutron_1993}
	Thompson C., Duncan R.~C., 1993, The Astrophysical Journal, 408, 194
	
	\bibitem[{Urpin {et~al}\mbox{.}(1986)Urpin, Levshakov, \&
		Iakovlev}]{urpin_generation_1986}
	Urpin V.~A., Levshakov S.~A., Iakovlev D.~G., 1986, Monthly Notices of the
	Royal Astronomical Society, 219, 703
	
	\bibitem[{{von Hoensbroech} \& {Xilouris}(1997)}]{1997A&A...324..981V}
	{von Hoensbroech} A., {Xilouris} K.~M., 1997, \aap, 324, 981
	
	\bibitem[{Woltjer(1964)}]{woltjer_x-rays_1964}
	Woltjer L., 1964, The Astrophysical Journal, 140, 1309
	
	\bibitem[{Wright(1973)}]{wright_pinch_1973}
	Wright G. A.~E., 1973, Monthly Notices of the Royal Astronomical Society, 162,
	339
	
	\bibitem[{{Young} \& {Rankin}(2012)}]{2012MNRAS.424.2477Y}
	{Young} S.~A.~E., {Rankin} J.~M., 2012, \mnras, 424, 2477
	
\end{thebibliography}
\end{document}